\documentclass[aps,prl,twocolumn,showpacs,superscriptaddress]{revtex4-1}
\usepackage{graphics}
\usepackage{graphicx}
\usepackage{dcolumn}
\usepackage{bm}
\usepackage{color}
\usepackage{amssymb,amsmath}
\UseRawInputEncoding

\begin{document}

\title{Ultra-broadband supercontinuum generation in gas-filled photonic-crystal fibers: The epsilon-near-zero regime}

\author{Mohammed F. Saleh and Fabio Biancalana}

\affiliation{SUPA, Institute of Photonics and Quantum Sciences, Heriot-Watt University, EH14 4AS Edinburgh, UK}

\begin{abstract}
In this Letter, we show theoretically that the nonlinear photoionisation process of a noble gas inside a hollow-core photonic crystal fibre can be exploited in obtaining broadband supercontinuum generation via pumping close to the mid-infrared regime. The interplay between the  Kerr and photoionisation nonlinearities is strongly enhanced in this regime.  Photoionisation continuously modifies the medium dispersion, in which the refractive index starts to significantly decrease and approach the epsilon-near-zero regime. Subsequently, the self-phase modulation induced by the Kerr effect is boosted because of the accompanied slow-light effect. As a result of this interplay, an output spectrum that comprises of a broadband light with multiple dispersive-wave emission is obtained.
\end{abstract}

\maketitle

\paragraph{Introduction---}
Supercontinuum generation is arguably one of the most extensively studied phenomenon in nonlinear waveguides \cite{Alfano70}. A short intense pulse develops into an almost flat octave-spanning spectrum as a consequence of  a multitude of nonlinear interactions between light and the hosting medium, such as self-phase modulation, self-steepening and shock formation, intrapulse Raman effect, solitonic propagation and fission, and dispersive wave generation from solitons \cite{Agrawal07}. In order to maximise the efficiency and the broadness of the supercontinuum, it is crucial to pump light near the zero of the anomalous group velocity dispersion (GVD) of the fiber, in the vicinity of which solitons (the main ingredient responsible for the spectral broadening) are more favourably formed. For this reason, special solid-core optical fibers with a microstructured cladding, known as photonic crystal fibers \cite{Knight96}, have been introduced in order to manipulate and shift the zero of the GVD closer to the desired laser-pump wavelength.

More recently, hollow-core photonic crystal fibers (HC-PCFs)  have opened the opportunity of investigating strong nonlinear light-gas interactions in structures with micrometer-square cross-sections and tens of centimetres long \cite{Travers11,Debord19}. Kagome HC-PCFs, in particular,  show  high confinement of light in the core with relatively low loss and low group velocity dispersion in the visible and near infrared regimes \cite{Russell03,Russell06}. Using these fibers, a series of interesting phenomena have been observed and predicted, such as stimulated Raman scattering at a very low pump energy \cite{Benabid02}, light-trapping molecules \cite{Alharbi16}, efficient deep-ultraviolet radiation \cite{Joly11}, soliton self-frequency blueshift \cite{Holzer11,Saleh11},  extremely unbalanced self-phase modulation \cite{Saleh12}, mid-infrared dispersive wave emission \cite{Novoa2015},  and ultra-broadband supercontinuum generation \cite{Belli15}.

A new class of media, known as epsilon-near-zero (ENZ) materials, have been recently utilised in obtaining a giant enhancement of Kerr nonlinearity via operating close to the plasma frequency $\omega_p$ \cite{Alam16,Caspani16,Kinsey19}.  In this Letter, we have exploited the photoionisation effect to approach the ENZ regime in gas-filled HC-PCFs. Via the self-compression effect, the peak intensities of micro-joule femtosecond pulses can easily reach the level sufficient to ionise the enclosed gas in a Kagome HC-PCF. Gases have refractive indices greater than but very close to one. However, due to the plasma created by the photoionisation effect, the medium effective refractive index can slightly, and sometimes considerably, drop below one. Roughly speaking, following the Drude model \cite{Saleh07}, the medium dielectric constant is proportional to $1-\omega_p^2/\omega^2$, where $\omega$ is the operating frequency. Hence, by working at longer wavelengths and increasing the pulse input energy to increase the amount of the generated plasma, we could significantly increase this reduction of the refractive index; allowing us to approach the ENZ regime in gas-filled fibers. 

As a result of operating in this new regime, we show that the fiber GVD can become dynamical for certain pump wavelengths near the mid-infrared range (MIR), and can be modulated by the effect of the presence of the plasma inside the core. This modulation occurs due to the fact that the linear refractive index felt by the pulse during propagation can be drastically lowered by the plasma dispersion, in such a way that the nonlinear refractive index of the gas can no longer be considered small when compared to the linear one. This allows to access a completely novel non-perturbative, ENZ regime in optical fibers, generating an ultra-broadband supercontinuum approximately from $0.5$ to $4$ $\mu$m together with multiple mid-infrared dispersive waves covering a range from $5$ to $10$ $\mu$m. This would open up exciting applicative possibilities in a variety of physics and biophysics areas, in particular in the field of molecular spectroscopy, where such broadband MIR sources are essential for accurate spectral analysis of complex biological tissues \cite{Kaya17}.

\paragraph{Modelling---}
Consider the propagation of a strong ultrashort pulse in a HC-PCF filled by a noble gas. The  spatial  evolution of the pulse electric field can be accurately determined via solving the unidirectional pulse-propagation equation (UPPE) \cite{Kolesik04,Kinsler11}
\begin{equation}
\dfrac{\partial\tilde{E}}{\partial z}=i\left [\beta\left( \omega\right)-\beta_{1}\omega \right] \tilde{E}+i\dfrac{\omega^{2}}{2c^{2}\epsilon_{0}\beta\left( \omega\right)}\tilde{P}_{\mathrm{NL}}, \label{Eq1}
\end{equation}
where $ z $ is the longitudinal propagation distance, $ \omega $ is the angular frequency,  $ \tilde{E}\left(z,\omega \right)  $ is the spectral electric field, $ \beta\left(\omega \right) $ is the combined gas and fiber dispersion of the fundamental mode,  $ \beta_{1}$ is the first-order dispersion coefficient,  $ c $ is the  speed of light in vacuum, $ \epsilon_{0} $ is the vacuum permittivity,  $\tilde{P}_{\mathrm{NL}} \left(z,\omega \right)=\mathcal{F}\left\lbrace P_\mathrm{NL}\left(z,t \right) \right\rbrace $ is the total nonlinear polarization, $\mathcal{F}$ is the Fourier transform, $ P_{\mathrm{NL}} = P_{\mathrm{K}}+P_{\mathrm{I}}$, $ P_{\mathrm{K}} =\epsilon_{0} \chi^{\left( 3\right)} E^{3}  $ is the Kerr nonlinearity,  $ \chi^{\left( 3\right) } $ is the third-order nonlinear susceptibility,  $ P_{\mathrm{I}}$ is the photoionisation-induced nonlinear polarisation, and $  t$ is the time in a reference frame moving with the pulse group velocity $ v_{g} =1/\beta_1$.

The photoionisation-induced polarisation $ P_{\mathrm{I}} $ at a position $z$  is governed by \cite{Geissler19} 
\begin{equation}
\dfrac{\partial P_{\mathrm{I}}}{\partial t}=\dfrac{e^{2}}{m_{e}}\displaystyle\int_{-\infty}^{t}n_{e}\left( z,t'\right)E\left(z,t'\right)dt'+\dfrac{\partial n_{e}}{\partial t}\dfrac{U_\mathrm{I}}{E\left(z,t\right)}, 
\label{Eq2}
\end{equation}
where $ e $,  $ m_{e} $  are the electron charge and mass, $ U_\mathrm{I} $ is the ionization energy of the gas, $ n_{e} $ is the generated free-electron density determined by
$ \partial_{t} n_{e}=\sigma\left(n_{\mathrm{T}}-n_{e} \right) $,  with $n_{ \mathrm{T} }$  the total density of the gas atoms, and  $  \sigma$  the ionization rate. Applying  the Ammosov-Delone-Krainov model \cite{Chang11}, which assumes tunnelling ionisation is dominant over multi-photon ionisation, 
\begin{equation}
\sigma\left(z,t\right)=\rho\left[\frac{\kappa}{\left|E\left(z,t\right)\right|}\right]^{2n-1}\, \exp\left[ \frac{ -\kappa}{3\left|E\left(z,t\right)\right|}\right] ,
\label{Eq3}
\end{equation}
where $ \rho = \frac{2^{2n}} {n\Gamma\left( n\right) \Gamma\left( n+1\right)} \frac{U _\mathrm{I}}{\hbar}  $,   $\kappa= \frac{4 U_\mathrm{I}\sqrt{2m_{e}U_{I}}}{\hbar e} $, $ n=\sqrt {\frac{U_\mathrm{H} }{U_\mathrm{I}}}$ is the effective principal quantum number, $ \Gamma $ is the gamma function,   $ U_\mathrm{H} =13.6$ eV is the atomic ionization energy of hydrogen, and $\hbar $ is the reduced Planck's constant. 

Equation (\ref{Eq2}) can be viewed as a superposition of plasma formation due to the liberation of electrons of the photoionisation process (the first term), and the associated losses with this process due to energy conversion (the second term). Substituting Eq. (\ref{Eq2}) in Eq. (\ref{Eq1}) results in 

\begin{equation}
\begin{array}{rl}
\dfrac{\partial\tilde{E}}{\partial z}=  &  i\left [\beta\left( \omega\right)-\beta_{1}\omega \right] \tilde{E}+i\dfrac{1}{2c^{2}\epsilon_{0}\beta\left( \omega\right)}\left[\epsilon_{0} \chi^{\left( 3\right)} \omega^{2}\mathcal{F}\left\lbrace E^{3}\right\rbrace \right.    \\
  & \left. - \frac{e^{2}}{m_{e}}\tilde{n}_{e}\otimes\tilde{E} + i\omega\,  \mathcal{F}
  \left\lbrace \frac{U_\mathrm{I}}{E} \sigma\left(n_{\mathrm{T}}-n_{e} \right)  \right\rbrace \right] ,  
\end{array}
\label{Eq4}
\end{equation}
with $\tilde{n}_{e}$ the spectral electron density and $\tilde{n}_{e}\otimes\tilde{E}=\int_{-\infty}^{\infty} \tilde{n}_{e}\left(z, \omega' \right)  \tilde{E}\left( z,\omega-\omega' \right) \, d\omega'$. In contrast to Raman nonlinearity that is described by a convolution between a response function and the pulse intensity in the time domain, plasma-induced nonlinearity is described by a convolution between the electron density and the electric field in the frequency domain. Using Taylor's series, the term $\tilde{E}\left( \omega-\omega' \right) $ can be expanded around $\omega$ as $ \tilde{E}\left( \omega\right)-\omega' \frac{\partial \tilde{E}}{\partial \omega} +\cdots$. Applying the zeroth-order approximation, $\tilde{n}_{e}\otimes \tilde{E} \approx \tilde{E}\left(z, \omega \right)\int_{-\infty}^{\infty} \tilde{n}_{e}\left( z,\omega' \right)   \, d\omega'={n}_{e}\left( z,t=0 \right)\tilde{E}\left(z, \omega \right)$.  Similarly, the self-phase modulation effect due to  Kerr nonlinearity and concurrent photoionisation loss can also be  approximated as $\approx E^2\left( z,0 \right) \tilde{E}\left(z, \omega \right)$ and $\frac{\omega \sigma U_\mathrm{I}}{E^2\left( z,0 \right) } \left[n_{\mathrm{T}}-n_{e}\left( z,0 \right)  \right]  \tilde{E}\left(z, \omega \right) $, respectively. In other words, these nonlinear contributions to the pulse propagation manifest themselves in modifying the propagation constant $ \beta\left(\omega \right)$ to be   
\begin{equation}
\begin{array}{l}
\beta_\mathrm{mod} \left( z, \omega\right)  =  \beta\left(\omega \right) \left(1+\Delta\beta_\mathrm{Kerr}+\Delta\beta_\mathrm{ion} + i \Delta\beta_\mathrm{loss}  \right),\vspace{1mm}\\ 
\Delta\beta_\mathrm{kerr}  \left( z, \omega\right) = \dfrac{\chi^{\left( 3\right)} \omega^{2}E^2\left( z,0 \right)}{2c^{2}\beta^2\left(  \omega\right)}, \\
\Delta\beta_\mathrm{ion}  \left( z, \omega\right) = - \dfrac{e^2 {n}_{e}\left( z,0 \right)}{2c^{2}m_e\epsilon_{0}\beta^2\left(  \omega\right)}= - \dfrac{\omega_p^2\left( z,0 \right)}{2c^{2}\beta^2\left(  \omega\right)}, \vspace{1mm} \\ 
\Delta\beta_\mathrm{loss} \left( z, \omega\right)  = \dfrac{\omega \sigma U_\mathrm{I}\left[n_{\mathrm{T}}-n_{e}\left( z,0 \right)  \right] }{2c^{2}\epsilon_{0}\beta^2\left(  \omega\right) E^2\left( z,0 \right) } ,
\end{array}
\label{Eq5}
\end{equation}
with  $\omega_p$ the dynamic plasma frequency. This  zeroth-order Taylor's approximation is valid for ultrashort pulses with broadband spectrum, and it has been numerically verified  versus the full convolution term. Whereas $ \Delta \beta_\mathrm{Kerr} $ has a weak dependence on the optical wavelength, $\Delta\beta_\mathrm{ion}  $ has a quadratic dependence on it. Therefore, operating at longer wavelengths would dramatically enhance the photoionisation-induced nonlinearity that results in decreasing the medium refractive index. Also, the associated  loss would  increase, however, linearly. Since the hosting nonlinear medium is a gas with a dielectric constant very close to unity, the plasma contribution could result in approaching the ENZ-regime, where the influence of Kerr nonlinearity on pulse propagation can be substantially enhanced due to the slow-light effect \cite{Kinsey19}.

\begin{figure}
\centerline{\includegraphics[width=8.6cm]{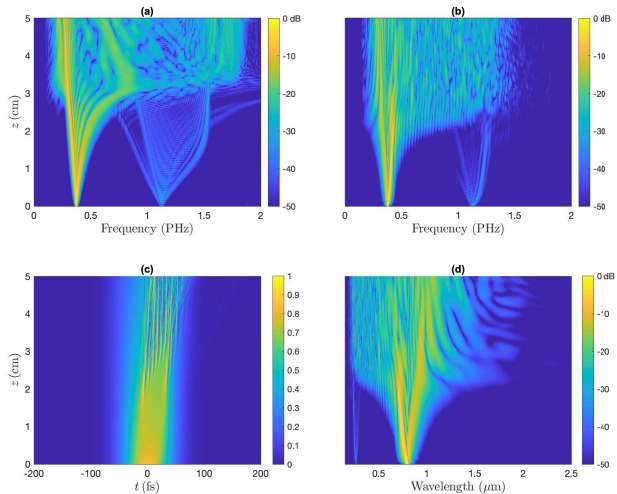}}
\caption{(Color online) (a,b) Spectral evolution of a Gaussian pulse with central wavelength $0.8\,\mu$m, energy $14\,\mu$J, and FWHM 50 fs in an argon-filled Kagome HC-PCF with a gas pressure 10 bar and a hexagonal core diameter $24\,\mu$m in the absence and presence of ionisation effects. (c,d) Temporal evolution of the pulse, and spectral evolution presented in the wavelength domain in the presence of ionisation effects. Contour plots in (a),(b), and (d) are given in a logarithmic scale and truncated at $-$50 dB.
\label{Fig1}}
\end{figure}

\paragraph{Simulations---}
The pulse dynamics of an intense ultrashort pulse with an energy of 14 $\mu$J and a central wavelength at 800 nm in an argon-filled photonic-crystal fiber is depicted in Fig. \ref{Fig1}. Panels (a,b) show the spectral evolution in the absence and presence of the photoionisation-induced nonlinearity, respectively. Photoionisation induces an earlier pulse compression as displayed in Fig. \ref{Fig1}(c), accompanied by the spectral broadening. However,  because of the associated  loss, the spectral broadening at the end is reduced by about $30\%$. Panel (d) is similar to (b) except it represents the spectral evolution in terms of the pulse wavelength to have a better visualisation of the spectrum in the MIR regime

\begin{figure}
\centerline{\includegraphics[width=8.6cm]{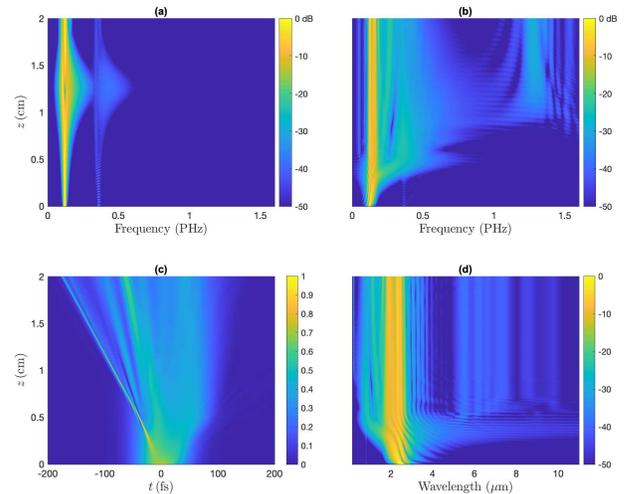}}
\caption{(Color online). (a,b) Spectral evolution of a Gaussian pulse with central wavelength $2.5\,\mu$m in an argon-filled HC-PCF  in absence and presence of ionisation effects. The rest of the parameters is the same as Fig. \ref{Fig1}.   (c,d) Temporal evolution of the pulse, and spectral evolution presented in the wavelength domain in the presence of ionisation effects.  The used set of simulation parameters are the same in the rest of the Letter, unless stated otherwise.
\label{Fig2}}
\end{figure}

Figure \ref{Fig2} repeats the simulations in Fig. \ref{Fig1},  except that the pulse central wavelength is moved to $2.5\, \mu$m in the MIR regime, which is experimentally accessible. As shown, a completely dramatic change in the pulse dynamics due to photoionisation takes place. In the absence of photoionisation,  Fig. \ref{Fig2}(a), there is a negligible  spectral broadening. However, by switching on the photoionisation nonlinearity, the pulse is massively broadened in the spectral domain from the deep-UV regime (the blue-side of the spectrum) to approximately  $10\,\mu$m in the MIR regime (the red-side). In contrast to operating close to the visible regime, photoionisation assists in this case in obtaining an ultra-broadband supercontinuum generation. As discussed in more details below, we attribute this to the change in the dispersion profile as well as the Kerr enhancement achieved by strong temporal localisation of the electric field, depicted in  Fig. \ref{Fig2}(c), associated with a significant reduction of the medium refractive index.

\begin{figure}
\centerline{\includegraphics[width=8.6cm]{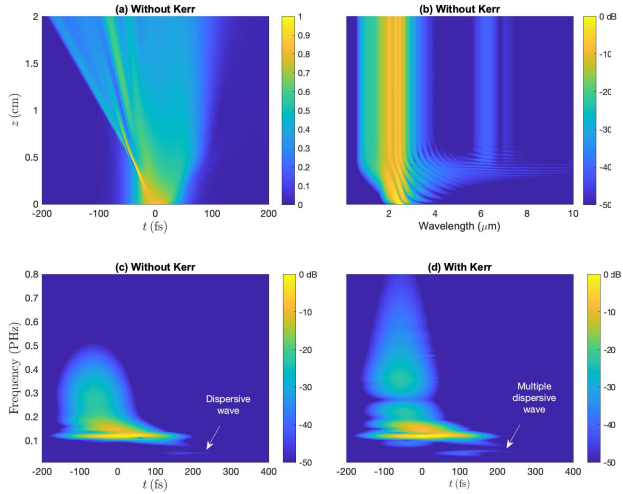}}
\caption{(Color online). (a,b) Temporal and spectral evolution in the absence of Kerr nonlinearity. (c,d) X-FROG spectrograms in the absence and presence of Kerr nonlinearity at $z=0.72$ cm. 
\label{Fig3}}
\end{figure}

To demonstrate that the photoionisation nonlinearity is the dominant factor when the central wavelength of the input pulse is in the MIR regime, we portray in Figs. \ref{Fig3}(a,b)  the pulse evolution in the  presence of only photoionisation effect by turning off the Kerr nonlinearity, since the influence of the latter  is minimal as suggested by Fig. \ref{Fig2}(a). As shown, the main feature of obtaining a broadband flat spectrum on the blue-side of the pulse at $ z\approx 0.4 $ cm, as well as the emission of few dispersive waves in the MIR regime between 6--7 $\mu$m still exist. 

To have deep insights in the physical processes involved in the pulse dynamics, we first display in Figs. \ref{Fig3}(c,d) the frames of cross frequency-resolved optical gating (X-FROG) spectrograms in the absence and presence of Kerr nonlinearity.  In the latter case, the pulse is significantly broadened,  blue-shifted to  more than its third-harmonic accompanied with emission of multiple dispersive waves on the other side of the spectrum. This suggests that the interplay between Kerr and photoionisation nonlinearities is being enhanced. in comparison with the cases when each one operates solely. 

\begin{figure}
\centerline{\includegraphics[width=8.6cm]{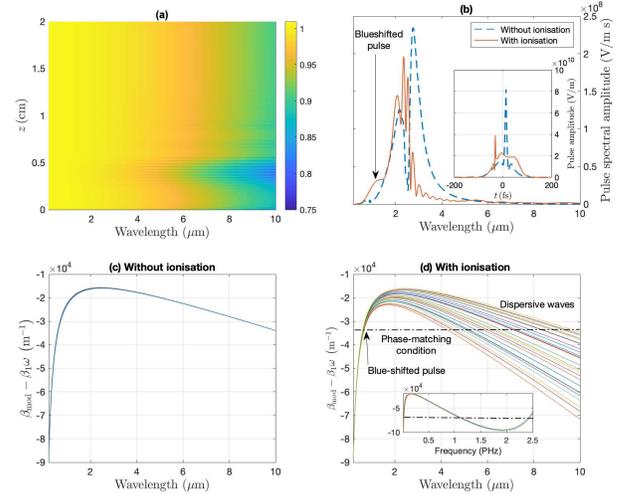}}
\caption{(Color online). (a) Evolution of the dielectric constant inside the fiber. (b) Comparison between the pulse spectrum in the absence and presence of the photoionisation process at the compression point. Inset compares the temporal profile of the pulse amplitude at the maximum compression point. (c,d) Phase accumulated per unit length in the absence and presence of photoionisation nonlinearity. Inset in (d) shows the possibility of dispersive wave emission in the deep-UV regime as well.
\label{Fig4}}
\end{figure}

In Fig. \ref{Fig4}(a), we depict the evolution of the real part of the effective dielectric constant, which can be determined via $\sqrt{\epsilon_r \left( z,\omega\right) }= c\beta_\mathrm{mod}  /\omega$  using Eq. (\ref{Eq5}), inside the fiber. Towards the end of the fiber, the dielectric constant varies from about 1 in the visible region to around 0.9 at longer wavelengths. However, at the position where the pulse compression and spectral broadening  take place ($ z \approx 0.4 $ cm),  the dielectric constant is  being reduced by more than 10$\%$  in comparison to the rest of the fiber.  This results in a decrease in the pulse group velocity $v_g$ that becomes very small as we approach the ENZ regime. Subsequently, the pulse electric field is boosted by a factor proportional to $1/v_g$  \cite{Kinsey19}, which in-turn results in a broader spectrum as the one displayed in Fig. \ref{Fig2}(d). Regarding the imaginary part of the dielectric constant induced by the concurrent ionisation loss, we have found that it is negligible in comparison to its real part.

The emission of a dispersive wave (DW) can be achieved via satisfying the phase matching condition with the blue-shifted pulse ($b$), i.e. $\phi\left(\omega_\mathrm{DW}\right)=\phi\left(\omega_b\right)$, with $\phi$ the accumulated phase. This condition turns to be $\beta_\mathrm{mod}\left( \omega_\mathrm{DW}\right)-\beta_{1}\omega_\mathrm{DW}=\beta_\mathrm{mod}\left( \omega_b\right)-\beta_{1}\omega_b$ via using Eqs. (\ref{Eq4},\ref{Eq5}).  Figure \ref{Fig4}(b) compares the pulse spectrum at the compression point in the absence and presence of the photoionisation process, whereas panels (c,d) depict the dependence of the parameter $\beta_\mathrm{mod}\left( \omega\right)-\beta_{1}\omega$ on the wavelength at different positions inside the fiber for both cases. In the absence of photoionisation, all the dispersion curves  approximately coincide. Moreover, the intensity of the blue-side of the pulse $\left(< 1\,\mu \mathrm{m}\right)$  is not sufficient enough to enhance any DW emission. On the other hand, photoionisation induces strong pulse compression with  sufficient energy on the  blue-side of the spectrum from 0.6 to 1 $\mu$m that can allow the emission of multiple dispersive waves in the MIR regime from 6 to 10 $\mu$m. These analytical predictions  show a very good agreement with the simulation results displayed in Fig. \ref{Fig2}(d), and also verify the approximations underpinned in Eq. (\ref{Eq5}).

\begin{figure}
\centerline{\includegraphics[width=8.6cm]{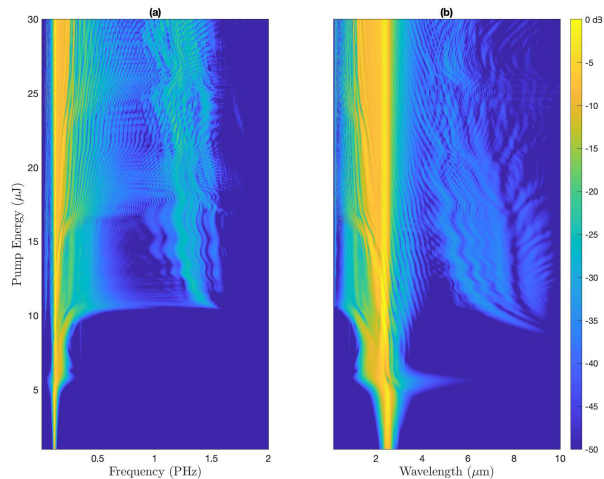}}
\caption{(Color online). Dependence of the spectral evolution on the pulse energy in the frequency and wavelength domains.
\label{Fig5}}
\end{figure}

Beside operating at longer wavelengths dramatically boost the effect of photoionisation nonlinearity, the input pulse energy is also an important  key parameter to access this new regime. Panels (a,b) in Fig. \ref{Fig5} depict the dependency of the pulse spectral evolution on its input energy in the frequency and wavelength domains, respectively.  As shown, there is a certain threshold for the input energy at which the pulse begins to exhibit this  ultra-broadband supercontinuum generation via emission of dispersive-waves in the UV and MIR regimes. This threshold represents the minimum energy at which the linear- and nonlinear-dispersion induced by plasma formation starts to be comparable. As the pulse energy increases, the MIR dispersive waves deteriorates at the far end because at this range of wavelengths the medium becomes opaque  or in other words the dielectric constant becomes negative  due to the plasma formation. However, the intensity of the blue-side of the spectrum is enhanced with increasing the pulse energy.

\paragraph{Conclusions---}
In this Letter, we have introduced  a new operating regime which exploits the nonlinear photoionisation process in gas-filled hollow-core photonic-crystal fibers. In this regime,  we launch intense femtosecond pulses with energies $>10\,\mu$J at the longest possible wavelength ($2.5 \, \mu$m) that can be experimentally accessible for this kind of laser sources. We showed how photoionisation assists in obtaining an ultra-broadband supercontinuum generation after a short propagation distance due to a plasma formation that (i) considerably changes the dispersion profile continuously inside the fiber, and (ii) enhances the Kerr nonlinearity of the system via localising the pulse electric field associated with a drop of the refractive index of the medium as the ENZ regime is approached. Moreover, we developed an analytical model based on the accurate UPPE to determine the Kerr and plasma contributions to the pulse propagation constant. The analytical predictions obtained by applying this model show a very good agreement with the numerical simulations. Finally, we believe that our results would open a new direction of research, where the photoionisation process would be beneficially exploited in developing new broadband light sources.

\section{Acknowledgement}
The authors thank Prof. J. Travers and Dr. Marcello Ferrera at Heriot-Watt University for useful discussions. This work is partially supported by M. Saleh's Royal Society of Edinburgh Research Fellowship.

 \bibliographystyle{apsrev4-1}	

%

\end{document}